\documentclass[aps,pre,twocolumn,epsf,graphicx,]{revtex4}
\usepackage{epsfig,latexsym}
\def\bec{\begin{center}}
\def\enc{\end{center}}

\def\ben{\begin{equation}}
\def\ba{\begin{array}}
\def\bea{\begin{eqnarray}}

\def\een{\end{equation}}
\def\eea{\end{eqnarray}}
\def\ea{\end{array}}
\def\btab{\begin{table}}
\def\btabu{\begin{tabular}}
\def\etab{\end{table}}
\def\etabu{\end{tabular}}
\def\bit{\begin{itemize}}
\def\eit{\end{itemize}}
\def\bef{\begin{figure}[htb]}
\def\befh{\begin{figure}[!h!]}
\def\enf{\end{figure}}

\def\L{\Lambda}

\def\b1{{\bf 1}}

\begin{document}

\title{\bf\noindent The thermal Casimir effect in lipid bilayer tubules} 

\author{D.S. Dean$^{(1,2)}$ and R.R. Horgan$^{(1)}$}

\affiliation{
(1) DAMTP, CMS, University of Cambridge, Cambridge, CB3 0WA, UK \\
(2) Laboratoire de Physique Th\'eorique,  UMR CNRS 5152, IRSAMC, Universit\'e 
Paul Sabatier, 118 route de Narbonne, 31062 Toulouse Cedex 04, France}
\date{25th October  2004}
\begin{abstract}
We calculate the thermal Casimir effect for a dielectric tube of
radius $R$ and thickness $\delta$ formed from a membrane in water. The
method uses a field-theoretic approach in the  grand canonical ensemble.
The leading contribution to the Casimir free energy behaves as 
$-k_BTL\kappa_C/R$ 
giving rise to an attractive force which tends to contract the 
tube. We find that $\kappa_C \sim 0.3$ for the case of typical lipid membrane
t-tubules. We conclude that except in the case of a very soft membrane
this force is insufficient to stabilize such tubes against the bending
stress which tends to increase the radius.
\end{abstract}
\maketitle
\vspace{.2cm} \pagenumbering{arabic} 
Lipid bilayers in water
exhibit a huge variety of geometries and in cell biology even more 
varied structures arise. In order to understand where biological mechanisms, 
such as molecular motors and
cytoskeletal structures, are determinant in the stability of biological
structures, one must first understand the role of the basic physical
interactions in pure model membrane systems.  
There has been much study of lipid bilayer
shape using standard continuum mechanics
\cite{helfa,boal}. This basic approach is also complemented by more
microscopic studies based on  lipid-lipid
interaction models \cite{boal,wurg}, this approach is of 
course ultimately necessary to
fully understand the physics of bilayers.  

In certain muscle cells, structures known as t-tubules are
found. These are basically cylindrical tubes whose surface is composed
of a lipid bilayer. Similar structures may also be mechanically drawn
off bilayer vesicles. The stability of these tubular structures
requires an explanation. The basic continuum theory 
(where the native curvature is zero) \cite{helfa,boal}
predicts that the free energy of a tube of length $L$ and radius $R$ is
\begin{equation}
F_B(L,R) = {k_B T\kappa_B L\over R}
\end{equation} 
where the above expression is strictly speaking the excess free energy
with respect to a flat membrane of the same area $A=2\pi RL$ and the
subscript $B$ refers to mechanical bending. Various experimental and
theoretical estimates for the bending rigidity
$\kappa_B$ can be found in the literature
and they lie between $1$ and $30$ (\cite{boal,wurg}).  The
values of $\kappa_B$ depend of course on the composition of the
bilayer and on the experimental protocol used to measure it. One
crucial element in both theoretical and experimental determinations of
$\kappa_B$ is whether the tube is attached to a reservoir of lipid or
not, {\em i.e.} whether the statistical ensemble is grand canonical or
grand canonical. Clearly if there is no reservoir then any increase in
the surface area of the tube will lead to a less dense lipid surface
concentration, in this case water may be able to become in contact
with the internal layer composed of the hydrophobic heads and so lead
to a significant increase in free energy. If upon changing the area of
the tube lipids can flow into the tube wall to maintain the local
optimal packing, then the free energy cost will be substantially
different.  This bending free energy is positive and hence in the absence
of external forces or extra constraints it tends to make a tube 
structure expand. We note that when lipid
tubes are drawn from a vesicle the mechanically applied tension can of
course overcome this free energy barrier. A natural question,
motivated by the fact that such structures occur in cells, is whether
other physical mechanisms  could lead to their formation and
explain their stability. A possible explanation put forward is that
that electrostatic effects involving surface
charges and ions (salt) in the surrounding medium could play a
role \cite{degen,wsalt,chjasi}. Certain experiments however showed a 
relative insensitivity of the equilibrium tube structure
to the concentration of  salt \cite{wsalt}. There are, however, 
systems with highly charged head groups where the salt concentration 
does appear important in determining the stability of the tubules
\cite{ssalt}. Another explanation has been put forward in terms of the
geometry of the lipid, notably the tail having a structure such that
there is a preferred orientation of the tails next to each other,
giving rise to a chirality which stabilizes the tubes
\cite{hepr,sesc,cen,fope,dejupr}. 

In this paper we investigate the possibility that a static van der
Waals or thermal Casimir force could provide an attractive force
across the tube leading to its stabilization. We adopt a continuum
model where the lipid bilayer is modeled as a layer of thickness
$\delta \approx 5-10$(nm) and of dielectric constant
$\epsilon_M\approx 2\epsilon_0$.  The surrounding water is also
treated as a dielectric continuum of dielectric constant $\epsilon_W
\approx 80 \epsilon_0$. We shall also adopt a model where the lipid
tube is fixed at each end to a flat lipid reservoir and so work in the
grand canonical ensemble. The behavior of systems composed of layers
of varying dielectric constants was first studied by Lifshitz and
coworkers \cite{lif,mani}. The formalism developed is an elegant way of
taking into account van der Waals forces in a continuum theory. Two
types of van der Waals forces are accounted for in theses theories,
firstly zero frequency van der Waals forces whose nature is classical
and secondly the frequency dependent ones due to temporal dipole
fluctuations. In terms of thermal field theory the former correspond
to modes with zero Matsubara frequency and the latter to the modes of
non-zero frequencies. These latter terms require information about the
frequency dependence of the dielectric constants, where as the former
only requires the static dielectric permittivity. In this paper we
will calculate the contribution of the  zero-frequency mode,
alternatively known as the thermal Casimir effect. The quantum 
Casimir effect corresponding to the modification of the ground state 
energy of the electromagnetic field has been intensively studied
in the case of idealized boundary conditions in a variety of
geometries including spheres and cylinders \cite{milton}. The thermal
Casimir effect investigated here has a similar mathematical structure
though the corresponding effective spatial dimension is one less. 
The temperature dependence of the full Casimir effect in a simplified 
model of a solid dielectric cylinder (and sphere) has been
recently examined using a heat kernel coefficient expansion \cite{bonepi}.  
In our analysis of  the diffuse limits we make use
of summation theorems for Bessel functions which were introduced for 
the study of the Casimir energy for cylinders with light-velocity 
conserving boundary conditions \cite{klro}.

We find that the thermal Casimir effect gives a contribution to the
excess free energy of the tube relative to that of the flat plane of
\begin{equation}
F_C(L,R) =
-\frac{k_B T L\kappa_C}{R}~+~O\left(\frac{\delta}{R^2}\right)~\ \label{eqdefkc},
\end{equation}
with
\begin{equation}
\kappa_C = \frac{\Delta^2}{64}\left[3\log(\L \delta) + 0.02954\right]
+\Delta^4B(\Delta^2) \label{eqkc}
\end{equation}
with $\Delta = (\epsilon_W-\epsilon_M)/(\epsilon_W+\epsilon_M)$
and where numerically we find that $B(\Delta^2)$ is only slowly varying
with $B(0) \sim 1/8$.  Here $\L \approx \pi/a$ is the large
wave-number, ultra-violet, cut-off which is phenomenologically
expressed in terms of microscopic cut-off $a$ corresponding to the
molecular size below which the continuum picture of the dielectric
medium breaks down. It is not a priori clear whether the cut off $a$
should be associated with the water or the lipid or indeed both; we
discuss this point later. In this expression the coefficient of
$\Delta^2$ is dominated by the logarithmic term, the constant
contribution being typically only of order 1\%, and the (estimated)
$O(\Delta^4)$ term is of comparable size for large enough $\Delta$.

We note that the sign of $F_C$ is negative and that it has the same
functional form as the bending free energy $F_B$, meaning that the
thermal Casimir force tends to collapse the tube and so helps to
stabilize the system against the bending forces. We shall show later
that with reasonable physical parameters $\kappa_C \approx
0.3-0.5$. We conclude that it is unlikely that the Casimir attraction
is able to overcome the repulsion due bending that is predicted by
current theories and measured by experiment.  However the results of
this letter are important for several reasons:
\begin{itemize}
\item We show that the thermal Casimir effect tends to contract the tube
structure.
\item The presence of the microscopic cut off in $\kappa_C$ shows that
the physics is ultimately dominated by the short scale or ultra-violet
physics. This means that weak electrolyte concentrations will have
little effect on the system as verified by experiments \cite{wsalt} given
that there are no strong surface charges.
\item Further attractive interactions will be
generated by the non-zero frequency Matsubara modes.
\item At a technical level use  the Pauli van Vleck formula
to evaluate the arising single body path integrals. This technique is
direct and well suited to layered systems.
\end{itemize}

In what follows we shall sketch the calculation of the thermal Casimir
free energy in the absence of electrolyte. The full calculation is
rather lengthy and it will be presented in the general case with
electrolyte in a more detailed longer paper \cite{inprep}.

The partition function for the zero frequency (static) fluctuations of
electrostatic field in a medium of varying dielectric constant
$\epsilon(\bf x)$ is  given by \cite{deho}
\begin{equation}
Z = \int d[\phi] \exp\left(-{\beta\over 2} \int d{\bf x} \epsilon(\bf
x) (\nabla \phi)^2\right). \label{eqfi}
\end{equation}  
In the model we shall consider the function $\epsilon({\bf x})=
\epsilon(r)$ where $r$ is the radial coordinate from the center of the
tube. The total length of the tube is
$L$ and the coordinate along the length of the tube is denoted by $z\in
[-L/2,L/2]$. At the extremities of the tube are two flat bilayers 
which act as a
reservoir for the material from which the tube is made. We consider
the limit $L\gg1$ so as to be able to neglect edge effects where the
tube meets the reservoir. The interior of the lipid tube is taken to
be at $R_1 = R-\delta/2$ and the exterior at $R_2=R+\delta/2$ In terms
of the coordinate $r$ we thus have
\begin{eqnarray}
\epsilon(r) &=& \epsilon_W\ \ r< R_1\ {\rm and} \ r > R_2\;, \nonumber
\\ \epsilon(r) &=& \epsilon_M \ \ R_1 < r < R_2\;.
\end{eqnarray}

Using the radial symmetry of the problem we express the field $\phi$
in cylindrical coordinates as
\begin{equation}
\phi(r,z,\theta) = \sum_{n= -\infty}^\infty\sum_{k} X_{n,k}(r)
\exp(ikz)\exp(in\theta)
\end{equation}
The functional integral now factorizes into a product of non-interacting 
path integrals. Taking the limit of $L$ large we find
\begin{equation}
F(R,\epsilon_W,\epsilon_M,\delta) = -k_BT L\sum_n \int {dk\over 2\pi}
\ln\left(Z_{n,k}(R,\epsilon_W,\epsilon_M,\delta)\right)\;,
\end{equation}
where
\begin{eqnarray}
& &Z_{n,k}(R,\epsilon_W,\epsilon_M,\delta)=
\int d[X]
\exp\left(-{\beta\over 2} \int_0^\infty \epsilon(r) \right.\nonumber \\
& &\left.\left[
\left(\frac{dX}{dr}\right)^2 + \left(\frac{n^2}{r^2}
+k^2\right)X^2\right]\right).
\end{eqnarray}

As we are only interested in the dependence of the free energy as a
function of $R$ we may work with the free energy normalized with respect to
that of  pure water with no tube, in this model it is equivalent to
subtracting off the free energy of a system where $\epsilon_W =
\epsilon_M$. This regularized free energy is thus
\begin{eqnarray}
&&F(R,\epsilon_W,\epsilon_M,\delta) = \nonumber \\ 
&-&k_BTL \sum_n \int {dk\over 2\pi}
\ln\left({Z_{n,k}(R,\epsilon_W,\epsilon_M,\delta)\over
Z_{n,k}(R,\epsilon_W,\epsilon_W,\delta)} \right).
\end{eqnarray}

In order to evaluate the individual kernels we make use of the
Pauli van Vleck formula which is exact for quadratic actions. If
$S[t_1,t_2,X]$ is an action quadratic in $X$, as is the case here, the
Pauli van Vleck formula gives
\begin{eqnarray}
&K&(t_1,t_2; X,Y)=\int_{X(t_1)=X}^{X(t_2)=Y} d[X] \exp(-S[t_1,t_2,X])
\nonumber \\ &=& \left(-{1\over 2\pi}{\partial^2 S[t_1,t_2,X_{cl}]\over
\partial X \partial Y}\right)^{1\over2} \exp(-S[t_1,t_2,X_{cl}]),
\end{eqnarray}
where $X_{cl}$ is the classical path minimizing the action
$S[t_1,t_2,X]$ given by $\delta S[t_1,t_2,X]/ \delta X = 0$ and with
the condition that the end points are fixed at $X$ and $Y$.  If the
action $S$ is quadratic then $S[t_1,t_2,X_{cl}])$ is a quadratic form
in $X$ and $Y$ and it only remains to evaluate the resulting
(ordinary) Gaussian integrals after obtaining the necessary
expressions for $S[t_1,t_2,X_{cl}])$, one obtains \cite{inprep}

\begin{eqnarray}
 {F(R,\epsilon_W,\epsilon_M,\delta)\over Lk_BT}&=& {1\over
R_1}g(\Lambda R_1,\Delta) + {1\over R_2}g(\Lambda R_2,-\Delta) \nonumber \\&+&
h(R_1,R_2,\Lambda,\Delta) + m(\Lambda,\Delta)\;, \label{Fghm}
\end{eqnarray} 
where $\Lambda = \pi/a$ is the ultra-violet cut-off
corresponding to the length scale $a$.  In Eq. (\ref{Fghm}) we have

\begin{eqnarray}
&&g(x,\Delta)= \frac{1}{2\pi}\sum_n\int_0^\Lambda
du\ln\left[1+ \Delta u (I_n(u)K_n(u))'\right]
\nonumber\\ 
&&h(R_1,R_2,\Lambda,\Delta)={1\over 2}\int_
0^{\Lambda } {dk \over \pi} \sum_n \ln\left[1 + \nonumber \right.\\ 
&&\left\{4\Delta^2 k^2
R_1R_2 I_n'(kR_1)I_n(kR_1) K'_n(kR_2) K_n(kR_2)\right./  \nonumber \\ 
&&\left(\left(1+ \Delta kR_1
(I_n(kR_1)K_n(kR_1))'\right) \times\right. \nonumber \\ 
&&\left.\left.\left(1- \Delta kR_2
(I_n(kR_2)K_n(kR_2))'\right)\right\}\right] \nonumber \\
&&m(\Lambda,\Delta)=-{1\over 2}\int_ 0^{\Lambda } {dk \over \pi}
\sum_n \ln(1-\Delta^2)\, \label{F}
\end{eqnarray}
where $I_n$ and $K_n$ are the modified Bessel functions
\cite{grad}. The contribution $Lk_BTg(\Lambda R,\Delta)/R$ is the free
energy of an isolated cylinder of length $L$, radius $R$ and
dielectric constant $\epsilon_M$ in a medium of dielectric constant
$\epsilon_W$. Thus the first two terms in Eq. (\ref{Fghm}) are the
respective separate contributions of the inner and outer cylindrical
regions that form the layer of thickness $\delta=R_2-R_1$; the term
$Lk_BTh(R_1,R_2,\Lambda,\Delta)$ is the contribution from the
interaction between the cylinders.  The function $g(x,\Delta)$
diverges as $x \to \infty$ and so this term in the free energy must be
regulated by taking a non-zero cut-off $a$. 
Viewed as a Taylor expansion in $\Delta$ we find that the
$O(\Delta)$ term of $g$ is independent of $r$ and thus in the free
energy the contributions proportional to $\Delta$ cancel. This to be
expected on physical grounds since by examining the limit of a diffuse
system one can see that any term proportional to $\Delta$ must be a
self energy term \cite{inprep}. The term of order $\Delta^2$ 
can be evaluated to be
\begin{equation}
g(x,\Delta)~=~-\frac{1}{256}\Delta^2\left[6\log(x) +30\log 2 +
6\gamma-11\right]~+~O(\Delta^4)\;.
\label{gD2}
\end{equation} 
This term is the leading term in the diffuse limit and the 
derivation uses Bessel function summation theorems exploited in
\cite{klro}. The same  form for the high temperature
expansion for a solid dielectric cylinder was obtained in \cite{bonepi}. 
 
The function $h$ is finite in this same limit and the term $m$, though
divergent, is independent of $R$ and cancels when the free energy of
the bulk in the reservoir is subtracted. The $O(\Delta^2)$
contribution to $h(R_1,R_2,\Lambda,\Delta)$ is given by 
\begin{eqnarray}
&&h(R-\delta/2,R+\delta/2,\Lambda,\Delta)=
\nonumber \\&&\frac{3}{64}\frac{\Delta^2}{R}\frac{1-y^2}{y^2}
\int_0^\infty
dz\frac{y^2z^4-1}{(1+z^2y^2)^{1/2}(1+z^2)^{5/2}},
\end{eqnarray}
where $y = \delta/2R$.

Taking account of the bulk reservoir and imposing overall conservation
of surface area, the relevant free energy is that of the tube less the
free energy of a flat membrane of the same area
\begin{equation}
F_C(R,\epsilon_W,\epsilon_M,\delta) =
F(R,\epsilon_W,\epsilon_M,\delta) -2\pi RL F_f\;,
\label{Fflat}
\end{equation}
where $F_f$ is the free energy per unit area of a flat membrane, which
is be given by
\begin{eqnarray}
F_f &=& {1\over 2\pi L}\lim_{R\to \infty}{1\over
R}F(R,\epsilon_W,\epsilon_M,\delta), \nonumber \\
&=& -{k_BT\over 16 \pi \delta^2}\sum_{m=1}^\infty {\Delta^{2m}\over m^3}
\end{eqnarray} 
and can be computed directly in a planar geometry by a variety of
methods \cite{mani}, including the path integral technique used here
\cite{deho}. The contribution to $F$ in this limit arises only from
the functions $h$ and $m$ in Eq. (\ref{Fghm}) and the latter cancels
identically. After subtraction, the dominant contribution,
$h_C(R,y,\Lambda,\Delta)$, from $h$ is the leading $1/R$ term, and we
find to $O(\Delta^2)$:
\begin{equation}
h_C(R,y,\Lambda,\Delta)~=~\frac{3}{64}\frac{\Delta^2}{R}\left(\log(y)+2\log
2 -1/2\right)\;.
\end{equation}

Putting our results together we find that $\kappa_C$, as defined by 
Eq. (\ref{eqdefkc}), is given by
\begin{eqnarray}
\kappa_C&=&
\frac{\Delta^2}{64}\left[3\log\left(\frac{\pi\delta}{a}\right)+6\log
2 + 3\gamma - 4\right]\nonumber \\&+& \Delta^4B(\Delta^2), \label{FC}
\end{eqnarray}
where the constant inside the bracket is evaluated to be
$0.02954\ldots$. An important point here is the exact cancellation
of the $\ln(R)$ terms coming from $g$ and $h$, giving a leading
order behavior of $F_C \sim 1/R$.  
The function $B(\Delta^2)$ receives contribution from
both the $g$ and $h$ terms in Eq. (\ref{F}) with $B(0) \ne
0$. Note that there are no odd terms in $\Delta$ in the leading $1/R$ behavior
of $F_C$ as, to leading order, one may set $\delta/R=0$  in the leading 
order behavior of $g$ and in the denominator of the
second term in the logarithm of the integral defining $h$.   
This is a consistent parametrization whilst $\delta \gg a$. The
limit $\delta \to 0$ must be taken carefully and when $\delta < a$ the
separation of $F$ in Eq. (\ref{F}) into contributions from functions
$g$ and $h$ is not useful since $h$ develops the compensating ultra-violet
divergence to that in $g$ and we find $\lim_{\delta \to 0}F_C = 0$, as
expected; in essence, the larger of $\delta$ and $a$ acts as the ultra-violet
cut-off on the integral defining $h$. We note that the dependence of 
$\kappa_C$ on the membrane thickness $\delta$ is considerably different
to that predicted for the mechanical bending energy, 
which is \cite{boal,rac} $\kappa_B\sim \delta^2$ and 
is verified experimentally \cite{rac} if an offset of the measured
lipid layer thickness $\delta$ is used. Of course $\kappa_B$ as
measured experimentally should be $\kappa_B +\kappa_C$ and the
form predicted here for $\kappa_C$ is compatible with the qualitative
behavior seen in measurements \cite{rac}.

In Table \ref{table}, for various values of $\Delta$ and $\delta/a$,
we compare the prediction of Eq. (\ref{FC}) with the result of
numerical integration and deduce a numerical value for $B(\Delta^2)$. Owing to
small systematic errors in the numerical calculation of the Bessel
functions there is a tiny discrepancy for very small $\Delta$ but
$B(\Delta^2)$ is seen to be a constant function from evaluations at
larger $\Delta$ and we see that $B(0)$ is plausibly $1/8$.

\btab \btabu{|c|c|c|c|c|}\hline
$\Delta$&$~~\delta/a$~~&\parbox{30mm}{$O(\Delta^2)$ coeff. of $1/R$\\
from Eq. (\ref{FC})}&\parbox{25mm}{Coeff. of $1/R$\\from
numerics}&$B(\Delta^2)$\\\hline
78/82&$10^3$&-0.342&-0.443&0.123\\\hline
78/82&$10^2$&-0.244&-0.346&0.123\\\hline
0.6&$10^3$&-0.1361&-0.1520&0.123\\\hline
0.6&$10^2$&-0.0972&-0.0162&0.123\\\hline
0.2&$10^3$&-0.0151&-0.0162&--\\\hline
0.6&$10^3$&-0.0038&-0.0040&--\\\hline \etabu
\caption{\label{table} For various values of $\Delta$ and $\delta/a$
we compare the prediction of Eq. (\ref{FC}) with numerical integration
and deduce a numerical value for $B(\Delta^2)$. Owing to
small systematic errors in the numerical calculation of the Bessel
functions there is a negligible discrepancy for very small $\Delta$
but $B(\Delta^2)$ is seen to be a constant function from evaluations
at larger $\Delta$. We see that the result for $F_C$ from
Eq. (\ref{FC}) is in very good agreement with the full
calculation. Various values of $\delta$ and $a$ were used but
typically $\delta = 1-10$(nm)} \etab

The physical value of the ultra-violet  cut-off length can only be determined
phenomenologically. This is because the model is an
effective field theory in which the dynamics of the molecular electric
dipoles is described by the dielectric constant which is a static
long-range parameter. The field modes with large-$k$ and $n$ probe the
static short distance properties of the model and so a more refined
field theory is needed for these scales. It is unclear whether the
molecular nature of the lipid has an effect on the ultra-violet cutoff but it
would seem most likely that the effective value of $\epsilon_W$ at
short scales ({\em i.e.} the microscopic details of water) are  dominant 
in this calculation. 

From Table \ref{table} we see that for a lipid bilayer tube in water
with $\delta=10$(nm) and $a=0.1$(nm) we find $\kappa_C = 0.346$. If we
were to include the contributions from the modes with non-zero
Matsubara frequencies, a calculation in progress, we can expected at
most a factor of two or so enhancement based on  past experience of similar
calculations (\cite{mani}), and so $\kappa_C \sim 1$ is a likely
largest value. These magnitudes are at the lowest end of those for $\kappa_B$
for known lipid bilayers in water (\cite{boal,wurg}). However, W\"urger
(\cite{wurg}) calculates $\kappa_B$ for surfactant films, analyzing the
role of hydrophobic tails, as a function of the tail length and the
area per molecule, and finds a wide range of values for $\kappa_B$
($\kappa_B = \pi \kappa$, with $\kappa$ from ref. (\cite{wurg}))
including values small enough, corresponding to soft interfaces, to
be balanced by our result. Thus it is conceivable that there can be small
tubes formed from soft membranes in water for which the bending forces
tending to expand the radius are compensated by the Casimir
attraction, and the tube is stabilized by sub-leading $O(1/R^2)$
forces.


\begin{thebibliography}{10}

\bibitem{helfa}
W.~Helfrich.
\newblock {\em Z. Naturforsch C}, 28:693, 1973.

\bibitem{boal}
D.~Boal.
\newblock Mechanics of the cell.
\newblock Cambridge University Press 2002.

\bibitem{wurg}
A.~W\"urger.
\newblock {\em Phys. Rev. Lett.}, 85:337, 2000.

\bibitem{degen}
P.G. de~Gennes.
\newblock {\em C.R. Acad. Sci. Paris}, 304:259, 1987.

\bibitem{wsalt}
J.S. Chappell and P.Yager.
\newblock {\em Biophys. J.}, 60:952, 1991.

\bibitem{chjasi}
M.V.~Jaric T.~Chou and E.D. Siggia.
\newblock {\em Biophysical J.}, 72:2042, 1997.

\bibitem{ssalt}
J.M.~Schnur M.A.~Marcowitz and A.~Singh.
\newblock {\em Chem. Phys. Lipids}, 62:193, 1992.

\bibitem{hepr}
W.~Helfrich and J.~Prost.
\newblock {\em Phys. Rev.}, A38:3065, 1988.

\bibitem{sesc}
J.V. Selinger and J.M. Schnur.
\newblock {\em Phys. Rev. Lett.}, 71:4091, 1993.

\bibitem{cen}
C.M. Chen.
\newblock {\em Phys. Rev.}, E59:6192, 1999.

\bibitem{fope}
J.B. Fournier and L.~Peliti.
\newblock {\em Phys. Rev. E}, E63:013901, 2000.

\bibitem{dejupr}
F.~Julicher I.~Derenyi and J.~Prost.
\newblock {\em Phys. Rev. Lett.}, 88:238101, 2002.

\bibitem{lif}
E.M.~Lifshitz I.E.~Dzyaloshinskii and L.P. Pitaevskii.
\newblock {\em Advan. Phys.}, 10:165, 1961.

\bibitem{mani}
J.~Mahantay and B.W. Ninhami.
\newblock Dispersion forces.
\newblock Academic Press 1976.

\bibitem{milton}
K.A. Milton.
\newblock {\em J. Phys.}, A37:R209, 2004.

\bibitem{bonepi}
V.V.~Nesterenko M.~Bordag and I.G. Pirozhenko.
\newblock {\em Phys. Rev.}, D65:045011, 2002.

\bibitem{klro}
I.~Klich and A.~Romeo.
\newblock {\em Phys. Lett.}, B476:369, 2000.

\bibitem{inprep}
D.S. Dean and R.R. Horgan.
\newblock in preparation.

\bibitem{deho}
D.S. Dean and R.R. Horgan.
\newblock {\em Phys. Rev.}, E65:061603, 2002.

\bibitem{grad}
I.S~Gradshteyn. et~al.
\newblock Table of integrals, series, and products.
\newblock Academic Press 2000.

\bibitem{rac}
W.~Rawicz et~al.
\newblock {\em Biophy. J.}, 79:328, 2000.

\end{thebibliography}
\end{document}